\PassOptionsToPackage{x11names,table,rgb,html}{xcolor}
\documentclass[british,screen,nonacm,final]{acmart}
\pdfoutput=1

\usepackage[T1]{fontenc}
\usepackage[utf8]{inputenc}
\usepackage{microtype}

\usepackage{xcolor}

\usepackage{babel}

\usepackage{amsmath}

\usepackage[inline]{enumitem}

\usepackage{listings}
\usepackage{lstcoq}
\definecolor{dkviolet}{rgb}{.5,0,.5}
\definecolor{dkblue}{rgb}{0,0,.5}
\definecolor{dkgreen}{rgb}{0,.5,0}
\definecolor{dkred}{rgb}{.5,0,0}
\definecolor{ltblue}{rgb}{0,.4,.6}

\usepackage{hyperref}
\usepackage{nameref}

\usepackage[capitalise,nameinlink,noabbrev]{cleveref}

\hypersetup{
	hidelinks,
	final,
}

\setcitestyle{numbers,sort&compress}
\setcopyright{none}

\newcommand{\NN}{\ensuremath{\mathbb N}}
\newcommand{\m}[1]{\ensuremath{\mathsf{#1}}}

\begin{document}

\title{Formalising a Turing-Complete Choreographic Language in Coq}

\author{Luís Cruz-Filipe}
\orcid{0000-0002-7866-7484}
\affiliation{
  \institution{University of Southern Denmark}
  \streetaddress{Campusvej 55}
  \city{Odense}
  \postcode{5230}
  \country{Denmark} 
}
\email{lcfilipe@gmail.com}

\author{Fabrizio Montesi}
\orcid{0000-0003-4666-901X}
\affiliation{
  \institution{University of Southern Denmark}
  \streetaddress{Campusvej 55}
  \city{Odense}
  \postcode{5230}
  \country{Denmark} 
}
\email{fmontesii@imada.sdu.dk}

\author{Marco Peressotti}
\orcid{0000-0002-0243-0480}
\affiliation{
  \institution{University of Southern Denmark}
  \streetaddress{Campusvej 55}
  \city{Odense}
  \postcode{5230}
  \country{Denmark} 
}
\email{peressotti@imada.sdu.dk}

\begin{abstract}
  Theory of choreographic languages typically includes a number of complex results that are
  proved by structural induction.
  The high number of cases and the subtle details in some of them lead to long reviewing
  processes, and occasionally to errors being found in published proofs.
  In this work, we take a published proof of Turing completeness of a choreographic language
  and formalise it in Coq. Our development includes formalising the choreographic language and its basic
  properties, Kleene's theory of partial recursive functions, the encoding of these functions as
  choreographies, and proving this encoding correct.
  
  With this effort, we show that theorem proving can be a very useful tool in the field of choreographic languages: besides the added degree of confidence that we get from a mechanised proof, the formalisation process led us to a significant simplification of the underlying theory.
  Our results offer a foundation for the future formal development of choreographic languages.
\end{abstract}




\maketitle

\section{Introduction}
\label{sec:introduction}

\paragraph*{Background}
In the setting of concurrent and distributed systems, choreographic languages are used to define interaction protocols that communicating processes should abide to~\cite{msc,bpmn,wscdl}. These languages are akin to the ``Alice and Bob'' notation found in security protocols, and inherit the key idea of making movement of data manifest in programs~\cite{NS78}. This is usually obtained through a linguistic primitive that looks like \lstinline+Alice.e -> Bob.x+, read ``\lstinline+Alice+ communicates the result of evaluating expression \lstinline+e+ to \lstinline+Bob+, which stores it in its local variable \lstinline+x+''.

In recent years, the communities of concurrency theory and programming languages have been prolific in developing methodologies based on choreographies, yielding results in program verification, monitoring, and program synthesis~\cite{Aetal16,Hetal16}. For example, in \emph{multiparty session types}, types are choreographies used for checking statically that a system of processes implements protocols correctly~\cite{HYC16}.
Further, in \emph{choreographic programming}, choreographic languages are elevated to full-fledged programming languages~\cite{M13p}, which can express how data should be pre- and post-processed by processes (encryption, validation, anonimisation, etc.). Choreographic programming languages showed promise in a number of contexts, including parallel algorithms~\cite{CM16}, cyber-physical systems~\cite{LNN16,LH17,GMP20}, self-adaptive systems~\cite{DGGLM17}, system integration~\cite{GLR18}, information flow~\cite{LN15}, and the implementation of security protocols~\cite{GMP20}.

\paragraph*{The Problem}
Proofs in the field of choreographic languages are extremely technical. They have to cover many cases, and they typically involve translations from/to other languages that come with their own structures and semantics.
The level of complexity makes peer-reviewing challenging. For example, it has recently been discovered that a significant number (at least $5$) of key results published in peer-reviewed articles on multiparty session types actually do not hold, and that their statements require modification~\cite{SY19}.

\paragraph*{This Article}
The aim of this article is to show that computer-aided verification -- in particular, interactive theorem proving -- can be successfully applied to the study of choreographies and to provide solid foundations for future developments.

Before presenting our scientific contributions, it is interesting to look at the story behind this article, as it tells us that interactive theorem proving is not just a tool to check what we already know.

Our development started in late 2018. Our starting point was the theory of Core Choreographies (CC), a minimalistic language that we previously proposed for the study of choreographic programming~\cite{CM20}.
CC is designed to include only the essential features of choreographic languages and minimal computational capabilities at processes (computing the successor of a natural number and deciding equality of two natural numbers). Nevertheless, it is expressive enough to be Turing complete, which is proven by developing a provably-correct translation of Kleene's partial recursive functions~\cite{Kleene52} into choreographies that implement the source functions by means of communication~\cite{CM20}.

At the TYPES conference in 2019, we gave an informal progress report on the formalisation of CC using the Coq theorem prover~\cite{CMP19}.
Our effort revealed soon a crux of unparsimonious complexity in the theory: a set of term-rewriting rules for a precongruence relation used in the semantics of the language for (i) expanding procedure calls with the bodies of their respective procedures and (ii) reshuffling independent communications in choreographies to represent correctly concurrent execution. This relation is closed under context and transitivity, and it can always be involved in the derivation of reductions, which led to tedious induction on the derivation of these term rewritings in almost all cases of proofs that had to do with the semantics of choreographies. In addition to being time consuming, formalising this aspect makes the theory presented in~\cite{CM20} much more complicated (we discuss this in \cref{discussion:precongr}).

\looseness=-1
At the time, the second author had been teaching a few editions of a course that includes theory of choreographies.
Interestingly, the same technical aspects that made the formalisation of CC much more intricate than its original theory were all found to be subtly complicated by students.
Motivated by this observation and our early efforts in formalising~\cite{CM20}, the teacher developed a revisited theory of CC for his course material that dispenses with the problematic notions and shows that they are actually \emph{unnecessary}~\cite{M20itc}. The choreography theory in~\cite{M20itc} is the one that we deal with in this article.

Thus, besides the scientific contributions that we are going to present, this article also shows that theorem proving can be used to do research: the insights that we got by doing this formalisation led to changes in the original theory. We are going to prove that this did not come at the cost of expressive power, by formalising that the same translation of Kleene's partial recursive functions from~\cite{CM20} still works as-is for the theory in~\cite{M20itc}. Furthermore, while formalising the theory we realised that some assumptions in some results were actually not necessary, which yielded stronger results.

\paragraph*{Contributions}
\looseness=-1
This article presents the first formalised theory of a full-fledged choreographic language, including its syntax and semantics, and the main properties of determinism, confluence, and deadlock-freedom by design. This theory is formalised in Coq, using its module system to make it parametric.
Furthermore, we formalise Kleene's theory of partial recursive functions (which, to the best of our knowledge, has not been done before). We then show that the choreographic language is Turing complete, by encoding these functions as choreographies and proving this encoding sound.

The full development can be downloaded at \cite{CMP21-source}.

\paragraph*{Structure}

We assume the reader to be familiar with Coq.
In \cref{sec:choreographies}, we present the syntax and semantics of our choreographic language, based on its formalisation in Coq, and establish the main theoretical properties of this language.
\cref{sec:kleene} presents the theory and formalisation of Kleene's partial recursive functions, and \cref{sec:turing} describes their encoding as choreographies and the proof of Turing completeness of the choreographic language.
We discuss the relevance of our results and future work in \cref{sec:discussion}.

\section{Choreographies}
\label{sec:choreographies}

In this section, we introduce the choreographic language of Core Choreographies (CC), together with the corresponding formalisation.

A choreography specifies a protocol involving different participants (processes) that can communicate among themselves and possess local computational capabilities.
Each process also has storage, which is accessible through variables.
There are two kinds of communications: value communications, where the sender process locally
evaluates an expression and sends the result to the receiver process, who stores it in one of its
variables; and label selection, where the sender selects one among some alternative behaviours (tagged by labels) offered by the receiver.
A choreography can also define (recursive) procedures, which can be invoked by their
name anywhere.
The formal syntax of choreographies is given in \cref{sec:chor-syntax}.

\subsection{Preliminaries}

We define the type of choreographies as a parametric Coq \lstinline+Module+, taking eight
parameters: the types of process identifiers (processes for short) \lstinline+Pid+, local variables \lstinline+Var+ (used to access
the processes' storage), values \lstinline+Val+, expressions \lstinline+Expr+, Boolean expressions
\lstinline+BExpr+, procedure names \lstinline+RecVar+ (from \emph{recursion variables}), and the
evaluation functions mapping expressions to values and Boolean expressions to Booleans.

The first six parameters are datatypes that are equipped with a decidable equality.
Due to difficulties with using the definitions in Coq's standard libraries, we reimplemented this as
a \lstinline+Module Type DecType+, and defined a functor \lstinline+DecidableType+ providing the
usual lemmas to simplify function definitions by case analysis on equality of two objects.

Evaluation requires a \emph{local state}, mapping process variables to actual values.
We model states as functors, taking \lstinline+Var+ and \lstinline+Val+ as parameters and returning a
\lstinline+Module+ containing this function type together with an operator to update the state
(by changing the value assigned to one variable) and lemmas characterising this operator.
For the semantics of choreographies, we also need \emph{global states}, which take \lstinline+Pid+
as an additional parameter and map each process to a local state.
This type is again enriched with operations to update a global state and their properties.

An evaluation function is a function mapping expressions to values, given a local state.
Evaluation must be compatible with extensional equality on states.
\begin{lstlisting}
Module Type Eval (Expression Vars Input Output : DecType).

Parameter eval : Expression.t -> (Vars.t -> Input.t) -> Output.t.
Parameter eval_wd : forall f f', (forall x, f x = f' x) -> forall e, eval e f = eval e f'.

End Eval.
\end{lstlisting}
This module is instantiated twice in the choregraphy type: with arguments \lstinline+Expr+,
\lstinline+Var+, \lstinline+Val+ and \lstinline+Val+, evaluating expressions to values, and with
arguments \lstinline+BExpr+, \lstinline+Var+, \lstinline+Val+ and \lstinline+bool+, evaluating
Boolean expressions to Booleans.

\subsection{Syntax}
\label{sec:chor-syntax}

We present the Coq definition of the type of choreographies, and afterwards briefly explain each
constructor and its pretty-printing.
\begin{lstlisting}
Inductive Label : Type :=
 | left : Label
 | right : Label.

Inductive Eta : Type :=
 | Com : Pid -> Expr -> Pid -> Var -> Eta
 | Sel : Pid -> Pid -> Label -> Eta.

Inductive Choreography : Type :=
 | Interaction : Eta -> Choreography -> Choreography
 | Cond : Pid -> BExpr -> Choreography -> Choreography -> Choreography
 | Call : RecVar -> Choreography
 | RT_Call : RecVar -> (list Pid) -> Choreography -> Choreography
 | End : Choreography.

Definition DefSet := RecVar -> (list Pid)*Choreography.

Record Program : Type :=
  { Procedures : DefSet;
    Main : Choreography }.
\end{lstlisting}
Constructor \lstinline+Interaction+ builds a choreography that starts with a communication
(\lstinline+Eta+), which can be either a value communication (\lstinline+Com+) or a label selection (\lstinline+Sel+).
These choreographies are written as \lstinline+p#e-->qdollarx;;C+ or \lstinline+p-->q[l];;C+,
respectively.
Label selections were inherited by choreographies from linear logic and behavioural types: they are used to communicate a choice made by the sender to the receiver.
In minimalistic theories of choreographies and behavioural types, it is common to restrict the set of labels that can be communicated (\lstinline+Label+) to two elements, generically called \lstinline+left+ and \lstinline+right+~\cite{CM20,CP10}.
These labels are typically used to propagate information about the local evaluation of a conditional expression, which generates two possible execution branches.

A choreography that starts by locally evaluating an expression is built using \lstinline+Cond+,
written \lstinline+If p??b Then C1 Else C2+, while invoking procedure \lstinline+X+ is built as
\lstinline+Call X+.
A procedure may involve several processes; the auxiliary term \lstinline+RT_Call X ps C+ represents
a procedure that has already started executing, but the processes in \lstinline+ps+ have not yet
entered it -- the term \lstinline+C+ is obtained from the actions of procedure that have been
executed (see the semantics below).
\lstinline+End+ denotes the terminated choreography.

A \lstinline+DefSet+ (set of procedure definitions) is a mapping assigning to each
\lstinline+RecVar+ a list of processes and a choreography; intuitively, the list of processes
contains the processes that are used the procedure.
A \lstinline+Program+ is a pair containing a set of procedure definitions and the choreography to
be executed at the start.

Terms built using \lstinline+RT_Call+ are meant to be runtime terms, generated while executing a
choreography; therefore, programs written by programmers should not contain such terms.
We call such a choreography \lstinline+initial+, and define it inductively in the natural way.

\paragraph*{Well-Formedness}
There are a number of well-formedness requirements on choreographies.
Some of these come from practical motivations and are typically explicitly required, while others
are more technical and not always written down in other articles.
We summarise these conditions.

A choreography is well-formed if its processes do not self-communicate: the two arguments of type
\lstinline+Pid+ to \lstinline+Com+ and \lstinline+Sel+ in all its subterms are always distinct.
Furthermore, the list of process names in the argument of \lstinline+RT_Call+ is never empty.
These conditions are defined separately by recursion in the natural way.

For a program \lstinline+P+ to be well-formed, there are more requirements on procedure
definitions and the annotations of the runtime terms.
First, both \lstinline+Main P+ and all choreographies in \lstinline+Procedures P+ must be
well-formed, and furthermore the latter must all be initial choreographies.
In \lstinline+Main P+, all runtime terms must be consistently annotated: the set \lstinline+ps+ in
\lstinline+RT_Call X ps C+ should only contain processes that occur in the definition of
\lstinline+X+ (as specified in \lstinline+Procedures P+).

Second, a program must be finitely specifiable.
Instead of requiring the type \lstinline+RecVar+ to be finite, which would significantly complicate
the formalisation, we require that there exist \lstinline+Xs:list RecVar+ such that every procedure
in \lstinline+Xs+, as well as \lstinline+Main P+, only calls procedures in \lstinline+Xs+.
Hence, it becomes irrelevant what the remaining procedure definitions are.
Well-formedness is thus parameterised on \lstinline+Xs+.

\begin{lstlisting}
Definition Program_WF (Xs:list RecVar) (P:Program) : Prop :=
  Choreography_WF (Main P) /\ within_Xs Xs (Main P) /\ consistent (Vars P) (Main P) /\
  forall X, In X Xs -> Choreography_WF (Procs P X) /\ initial (Procs P X) /\ (Vars P X) <> nil /\ within_Xs Xs (Procs P X).
\end{lstlisting}

The third and last condition is that the set of procedure definitions in a program must be well-annotated: if
\lstinline+Procedures P X=(ps,C)+, then the set of processes used in \lstinline+C+ must be included
in \lstinline+ps+.
The set of processes used in \lstinline+C+ is in turn recursively defined using the information in
\lstinline+Procedures P+, so computing an annotation is not straightforward.
For this reason, it can be convenient in practice to over-annotate a program -- which is why
well-annotation only requires a set inclusion.
\begin{lstlisting}
Definition well_ann (P:Program) : Prop :=
  forall X, set_incl _ (CCC_pn (Procs P X) (Vars P)) (Vars P X).

Definition CCP_WF (P:Program) := well_ann P /\ exists Xs, Program_WF Xs P.
\end{lstlisting}
While \lstinline+Program_WF+ is decidable, \lstinline+well_ann+ and \lstinline+CCP_WF+ (for CC
program) are not, due to the quantifications in their definitions.
This motivates defining \lstinline+Program_WF+ separately.

Formalising well-formedness requires some auxiliary definitions (the sets of processes and procedure
names used in a choreography) and several inductive definitions.
Most of them are straightforward, if sometimes cumbersome; the complexity of the final definition
can make proofs of well-formedness quickly grow in size and number of cases, so we provide a number
of inversion results such as the following to make subsequent proofs easier.
\begin{lstlisting}
Lemma CCP_WF_eta : forall Defs C eta,
  CCP_WF (Build_Program Defs (eta;;C)) -> CCP_WF (Build_Program Defs C).
\end{lstlisting}

\subsection{Semantics}

The semantics of CC is defined by means of labelled transition systems, in several layers.
At the lowest layer, we define the transitions that a choreography can make (\lstinline+CCC_To+),
parameterised on a set of procedure definitions; then we pack these transitions into the more usual
presentation -- as a labelled relation \lstinline+CCP_To+ on \emph{configurations} (pairs
program/state).
Finally, we define multi-step transitions \lstinline+CCP_ToStar+ as the transitive and reflexive
closure of the transition relation.
This layered approach makes proofs about transitions cleaner, since the different levels of induction
are separated.

\paragraph*{Transition Labels}
We have two types of transition labels.
The first one is a simple inductive type with constructors corresponding to the possible actions a
choreography can take: value communications, label selections, local conditional, or local procedure
call.
This type is called \lstinline+RichLabel+: rich labels are not present in the informal theory~\cite{M20itc}, but they are needed to obtain strong enough induction hypotheses in proofs of results about CC that we will need in \cref{sec:turing} for Turing completeness.
The labels in the informal theory correspond to observable actions; they are formalised as
\lstinline+TransitionLabel+, and they forget the internal details of actions.
The two types are connected by a function \lstinline+forget:RichLabel -> TransitionLabel+.
\begin{lstlisting}
Inductive RichLabel : Type :=
| R_Com (p:Pid) (v:Value) (q:Pid) (x:Var) : RichLabel
| R_Sel (p:Pid) (q:Pid) (l:Label) : RichLabel
| R_Cond (p:Pid) : RichLabel
| R_Call (X:RecVar) (p:Pid) : RichLabel.

Inductive TransitionLabel : Type :=
| L_Com (p:Pid) (v:Value) (q:Pid) : TransitionLabel
| L_Sel (p:Pid) (q:Pid) (l:Label) : TransitionLabel
| L_Tau (p:Pid) : TransitionLabel.
\end{lstlisting}

\paragraph*{Transition Relations}
\lstinline+CCC_To+ is defined inductively by a total of 11 clauses, corresponding to the 11 rules in the informal presentation.
We include some of them below, with some proof terms omitted.
\begin{lstlisting}
Inductive CCC_To (Defs : DefSet) :
  Choreography -> State -> RichLabel -> Choreography -> State -> Prop :=
 | C_Com p e q x C s s' : let v := (eval_on_state e s p) in
        eq_state_ext s' (update s q x v) ->
        CCC_To Defs (p # e --> qdollarx;; C) s (R_Com p v q x) C s'
 | C_Delay_Eta eta C C' s s' t: disjoint_eta_rl eta t -> 
        CCC_To Defs C s t C' s' ->
        CCC_To Defs (eta;; C) s t (eta;; C') s'
 | C_Call_Start p X s s':
        eq_state_ext s s' ->
        set_size _ (fst (Defs X)) > 1 -> In p (fst (Defs X)) ->
        CCC_To Defs
               (Call X) s
               (R_Call X p)
               (RT_Call X (set_remove _ p (fst (Defs X))) (snd (Defs X))) s'
 | C_Call_Enter p ps X C s s':
        eq_state_ext s s' -> set_size _ ps > 1 -> In p ps ->
        CCC_To Defs
               (RT_Call X ps C) s
               (R_Call X p)
               (RT_Call X (set_remove _ p ps) C) s'
 | C_Call_Finish p ps X C s s':
        eq_state_ext s s' -> set_size _ ps = 1 -> In p ps ->
        CCC_To Defs
               (RT_Call X ps C) s (R_Call X p) C s'.
\end{lstlisting}
The first constructor defines a transition for a value communication, with the state being updated with the value received at the receiver.
Since states are functions, we do not want to require that the resulting state be
\lstinline+update s q x v+ -- the state obtained directly from \lstinline+s+ by updating the value
at \lstinline+q+'s variable \lstinline+x+ with \lstinline+v+ -- but only that it be extensionally
equal to it (the values of variables at all processes are the same).
(The stronger requirement would break, e.g., confluence, since updating two different variables in a different order does not yield the same state.)

The original informal theory allows for out-of-order execution of independent interactions~\cite{CM20,M20itc}, a well-established feature of choreographic languages~\cite{CM13,HYC16}. For example, given a choreography that consists of two independent communications such as \lstinline+p#e-->qdollarx;;r#e'-->sdollary;;End+ (``\lstinline+p+ communicates \lstinline+e+ to \lstinline+q+ and \lstinline+r+ communicates \lstinline+e'+ to \lstinline+s+'') where \lstinline+p+, \lstinline+q+, \lstinline+r+, and \lstinline+s+ are distinct processes, we should be able to observe the first and the second value communication in whichever order.
Out-of-order execution is modelled by three rules, of which the second constructor shown is an example.
Here, a choreography is allowed to reduce under a prefix \lstinline+eta+ if its label does not share
any processes with \lstinline+eta+.
This side-condition is checked by \lstinline+disjoint_eta_rl eta t+; several auxiliary predicates
named \lstinline+disjoint_type1_type2+ are defined to simplify writing these conditions.

Procedure calls are managed by four rules, of which the main three are shown.
A procedure call is expanded when the first process involved in it enters it (rule \lstinline+C_Call_Start+).
The remaining processes and the procedure's definition are stored in a runtime term, from which we can observe transitions either by more processes entering the procedure (rule \lstinline+C_Call_Enter+) or by
out-of-order execution of internal transitions of the procedure (rule \lstinline+C_Delay_Call+, not
shown).
When the last process enters the procedure, the runtime term is consumed (rule
\lstinline+C_Call_Finish+).
The missing rule addresses the edge case when a procedure only uses one process.

In order to prove results about transitions, it is often useful to infer the resulting choreography and state. The constructors of \lstinline+CCC_To+ cannot be used for this purpose, since the resulting state is
not uniquely determined.
Therefore, we prove a number of lemmas stating restricted forms of transitions that are useful for
forward reasoning.
\begin{lstlisting}
Lemma C_Com' : forall Defs p e q x C s, let v := (eval_on_state e s p) in
        CCC_To Defs (p # e --> qdollarx;; C) s (R_Com p v q x) C (update s q x v).
\end{lstlisting}

Afterwards, we formalise the transition relations as defined in~\cite{M20itc}.
\begin{lstlisting}
Definition Configuration : Type := Program * State.

Inductive CCP_To : Configuration -> TransitionLabel -> Configuration -> Prop :=
 | CCP_To_intro Defs C s t C' s' : CCC_To Defs C s t C' s' ->
     CCP_To (Build_Program Defs C,s) (forget t) (Build_Program Defs C',s').

Inductive CCP_ToStar : Configuration -> list TransitionLabel -> Configuration -> Prop :=
 | CCT_Refl c : CCP_ToStar c nil c
 | CCT_Step c1 t c2 l c3 : CCP_To c1 t c2 -> CCP_ToStar c2 l c3 -> CCP_ToStar c1 (t::l) c3.
\end{lstlisting}
We also define the suggestive notations \lstinline+c --[ tl ]--> c'+ for \lstinline+CCP_To c tl c'+
and \lstinline+c --[ ts ]-->* c'+ for \lstinline+CCP_To_Star c ts c'+.

\subsection{Progress, Determinism, and Confluence}

The challenging -- and interesting -- part of formalising CC is establishing the basic properties of
the language, which are essential for more advanced results and typically not proven in detail in
publications. We discuss some of the issues encountered, as these were the driving force behind the changes relative to~\cite{CM20}.

The first key property of choreographies is that they are deadlock-free by design: any choreography
that is not terminated can reduce.
Since the only terminated choreography in CC is \lstinline+End+, this property also implies that any
choreography either eventually reaches the terminated choreography \lstinline+End+ or runs infinitely.
These properties depend on the basic result that transitions preserve well-formedness.
\begin{lstlisting}
Lemma CCC_ToStar_CCP_WF : forall P s l P' s', CCP_WF P -> (P,s) --[l]-->* (P',s') -> CCP_WF P'.

Theorem progress : forall P, Main P <> End -> CCP_WF P -> forall s, exists tl c', (P,s) --[tl]--> c'.

Theorem deadlock_freedom : forall P, CCP_WF P ->
  forall s ts c', (P,s) --[ts]-->* c' -> {Main (fst c') = End} + {exists tl c'', c' --[tl]--> c''}.
\end{lstlisting}

This is the first place where we benefit from the change in both the syntax and semantics of choreographies from~\cite{CM20} to~\cite{M20itc}, which removes idiosyncrasies that required clarifications in the reviewing process of~\cite{CM20}.
In the language of~\cite{CM20}, procedures are defined inside choreographies by means of a $\m{def}\ X=C_X\ \m{in}\ C$ constructor in choreographies.
This makes the definition of terminated choreography much more complicated, since \lstinline+End+ could occur inside some of these terms.
Furthermore, procedure calls were expanded by structural precongruence, so that a choreography as $\m{def}\ X=\m{End}\ \m{in}\ X$ would also be terminated.
Separating procedure definitions from the main choreography in a program and promoting procedure calls to transitions makes stating and proving progress much simpler.
The direct syntactic characterisation of termination also has advantages, since it is intuitive and easily verifiable.

The second key property is confluence, which is an essential ingredient of the proof of Turing completeness below: if a choreography has two different transition paths, then these paths either end at the same configuration, or both resulting configurations can reach the same one.
This is proved by first showing the diamond property for choreography
transitions, then lifting it to one-step transitions, and finally
applying induction.

\begin{lstlisting}
Lemma diamond_Chor : forall Defs C s tl1 tl2 C1 C2 s1 s2,
  CCC_To Defs C s tl1 C1 s1 -> CCC_To Defs C s tl2 C2 s2 ->
  tl1 <> tl2 -> exists C' s', CCC_To Defs C1 s1 tl2 C' s' /\ CCC_To Defs C2 s2 tl1 C' s'.

Lemma diamond_1 : forall c tl1 tl2 c1 c2,
  c --[ tl1 ]--> c1 -> c --[ tl2 ]--> c2 ->
  tl1 <> tl2 -> exists c', c1 --[ tl2 ]--> c' /\ c2 --[ tl1 ]--> c'.

Lemma diamond_4 : forall P s tl1 tl2 P1 s1 P2 s2,
  (P,s) --[ tl1 ]-->* (P1,s1) -> (P,s) --[ tl2 ]-->* (P2,s2) ->
  (exists P' tl1' tl2' s1' s2', (P1,s1) --[ tl1' ]-->* (P',s1') /\ (P2,s2) --[ tl2' ]-->* (P',s2') /\ eq_state_ext s1' s2').
\end{lstlisting}

As an important consequence, we get that any two executions of a choreography that end in a terminated
choreography must yield the same state.
\begin{lstlisting}
Lemma termination_unique : forall c tl1 c1 tl2 c2,
  c --[tl1]-->* c1 -> c --[tl2]-->* c2 -> Main (fst c1) = End -> Main (fst c2) = End -> eq_state_ext (snd c1) (snd c2).
\end{lstlisting}

\label{discussion:precongr}
The complexity of the proof of confluence was the determining factor for deciding to start our work from the variation of the choreographic language presented in~\cite{M20itc} instead of that in~\cite{CM20}.
The current proof of confluence takes about 300 lines of Coq code, including a total of 11 lemmas.
This is in stark contrast with the previous attempt, which already included over 30 lemmas with
extremely long proofs.
The reason for this complexity lay, again, in both inlined procedure definitions and structural precongruence.
Inlined procedure definitions forced us to deal with all the usual problems regarding bound variables and renaming; structural precongruence introduced an absurd level of complexity because it allowed choreographies to be rewritten arbitrarily.

To understand this issue, consider again the choreography \lstinline+p#e-->qdollarx;; r#e'-->sdollary;; End+.
As we have previously discussed, this choreography can execute first either the communication between \lstinline+p+ and \lstinline+q+ or the one between
\lstinline+r+ and \lstinline+s+.
In our framework, the first transition is modelled by rule \lstinline+C_Com+, while the second is obtained by applying rule \lstinline+Delay_Com+ followed by \lstinline+C_Com+.
In a framework with reductions and structural precongruence, instead, the second transition is modelled by first rewriting the choreography as \lstinline+r#e'-->sdollary;; p#e-->qdollarx;; End+ and then applying
rule \lstinline+C_Com+~\cite{CM20}.
The set of legal rewritings is formally defined by the structural precongruence relation $\preceq$,
and there is a rule in the semantics allowing $C_1$ to reduce to $C_2$ whenever $C_1\preceq C'_1$,
$C'_2\preceq C_2$, and $C'_1$ reduces to $C_2$.
Thus, the proof of confluence also needs to take into account all the possible ways into which choreographies may be rewritten in a reduction.
In a proof of confluence, where there are two reductions, there are four possible places where
choreographies are rewritten; given the high number of rules defining structural precongruence, this
led to an explosion of the number of cases.
Furthermore, induction hypotheses typically were not strong enough, and we were forced to resort to complicated auxiliary notions such as explicitly measuring the size of the derivation of transitions, and proving that
transitions could be normalised.
This process led to a seemingly ever-growing number of auxiliary lemmas that needed to be proved, and after several months of work with little progress it became evident that the problem lay in the formalism.

With the current definitions, the theory of CC is formalised in two files.
The first file, which defines the preliminaries, contains 40 definitions, 58 lemmas and around 700 lines of code. The second file, which defines CC-specific results, contains 48 definitions, 106 lemmas, 2 theorems and around 2100 lines of code.

\section{Partial Recursive Functions}
\label{sec:kleene}

In order to formalise Turing completeness of our choreographic language, we need a model of
computation.
In~\cite{CM20}, the model chosen was Kleene's partial recursive functions~\cite{Kleene52}, and the
proof proceeds by showing that these can all be implemented as a choreography, for a suitable
definition of implementation.
This proof structure closely follows that of the original proof of computational completeness for
Turing machines~\cite{Turing36}.

In this section, we describe our formalisation of partial recursive functions, and the main challenges
and design options that it involved.
Following standard pratice, we routinely use lambda notation for denoting these functions.

\subsection{Syntax}
The class of partial recursive functions is defined inductively as the smallest class containing the
constant unary zero function $Z=\lambda x.0$, the unary successor function $S=\lambda x.x+1$ and the
$n$-ary projection functions $P^m_k=\lambda x_1\ldots x_m.x_k$ (base functions), and closed under
the operations of composition, primitive recursion, and minimisation.
All functions have an arity (natural number); the arity of $Z$ and $S$ is $1$, and the arity of
$P^m_k$ is $m$.
Given a function $g$ of arity $m$ and $m$ functions $f_1,\ldots,f_m$ of arity $k$, then the
composition $C(g,\vec f)$ has arity $k$; if $g$ has arity $k$ and $h$ has arity $k+2$, then function
$R(g,h)$ defined by primitive recursion from $g$ and $h$ has arity $k+1$; and if $h$ has arity
$k+1$, then its minimisation $M(h)$ has arity $k$.

We formalise this class as a dependent inductive type \lstinline+PRFunction+ taking the arity of the
function as a parameter.
In order to ensure the correct number of arguments in composition, we require $f_1,\ldots,f_m$ to be
given as a vector of length $m$.

\begin{lstlisting}
Inductive PRFunction : nat -> Set :=
  | Zero : PRFunction 1
  | Successor : PRFunction 1
  | Projection : forall {m k:nat}, k < m -> PRFunction m
  | Composition : forall {k m:nat} (g:PRFunction m) (fs:t (PRFunction k) m), PRFunction k
  | Recursion : forall {k:nat} (g:PRFunction k) (h:PRFunction (2+k)), PRFunction (1+k)
  | Minimization : forall {k:nat} (h:PRFunction (1+k)), PRFunction k.
\end{lstlisting}
Note the required proof term on the constructor for projections.
The parameter \lstinline+k+ is one unit lower than the parameter $k$ in the mathematical definition,
since Coq's natural numbers start at $0$ -- this choice simplifies the development.

\subsection{Semantics}

A partial recursive function of arity $m$ is meant to denote a partial function of type
$\NN^m\to\NN$.
The denotation of $Z$, $S$ and $P^m_k$ was already given above; the remaining operators are
interpreted as follows, where we write $\vec x$ for $x_1,\ldots,x_k$.
\begin{align*}
  C(g,\vec f)(\vec x) &= g(f_1(\vec x),\ldots,f_m(\vec x)) \\
  R(g,h)(0,\vec x) &= g(\vec x) \\
  R(g,h)(n+1,\vec x) &= h(n,R(g,h)(n,\vec x),\vec x) \\
  M(h)(\vec x) &=n\mbox{ if $h(\vec x,n)=0$ and $h(\vec x,i)>0$ for all $0\leq i<n$}
\end{align*}

Minimization can lead to partiality, since there may be no $n$ satisfying the conditions given in
its definition.
This partiality propagates, since any value depending on an undefined value is also undefined.
Kleene's original work does not completely specify this mechanism: for example, if $f$ is a unary
function that is undefined everywhere, should $C(Z,f)$ be also everywhere undefined, or constantly
zero?
It is generally assumed that a function is undefined whenever any of its arguments is undefined,
even in the case where those arguments are not used for computing the result; we take this approach
in this work.

Since Coq does not allow for defining partial functions, we take an operational approach to
the semantics of partial recursive functions, and interactively define the bounded evaluation of an
$m$-ary function $f$ on a vector of length $m$ in $n$ steps, which has type \lstinline+option nat+.
This construction proceeds in three steps.
First, we deal with minimisation by defining
\begin{lstlisting}
Fixpoint find_zero_from {k} (h:t (option nat) (1+k) -> option nat)
  (ns:t (option nat) k) (init:nat) (steps:nat) : option nat :=
  match steps with
  | O => None
  | S m => match h (shiftin (Some init) ns) with
           | None => None
           | Some O => Some init
           | Some (S _) => find_zero_from h ns (S init) m end end.
\end{lstlisting}
which tries to find the smallest zero of \lstinline+h+ starting at \lstinline+init+ using a bound of
\lstinline+steps+ steps for the first value, \lstinline+steps-1+ for the next value, etc.
(The type \lstinline+t T n+ is the type of vectors containing exactly \lstinline+n+ elements of type
\lstinline+T+.)
With this, we recursively define the evaluation function
\begin{lstlisting}
Fixpoint eval_opt {m} (f:PRFunction m) : forall (steps:nat) (ns:t (option nat) m), option nat.
\end{lstlisting}
Given the complexity of \lstinline+PRFunction+, this function is defined interactively.
Finally, we define
\begin{lstlisting}
Definition eval {m} (f:PRFunction m) (steps:nat) (ns:t nat m) : option nat
  := eval_opt f steps (map Some ns)
\end{lstlisting}
as our intended evaluation function.

Evaluation starts by checking that all arguments are defined.
If this is the case, then the base functions always return their value; composition and recursion
call the functions that they depend upon with the same number of steps; and minimisation initiates a
search from $0$ with the bounds explained above.

In order to ensure that the interactive definitions are correct, we prove a number of lemmas stating
that the defining equations for each class of functions hold.
For example, for recursion we have the following three lemmas.
\begin{lstlisting}
Lemma Recursion_correct_base : forall k (g:PRFunction k) (h:PRFunction (2+k)) (ns:t nat (1+k)),
  forall steps, hd ns = 0 -> eval (Recursion g h) steps ns = eval g steps (tl ns).

Lemma Recursion_correct_step : forall k (g:PRFunction k) (h:PRFunction (2+k)) (ns:t nat (1+k)),
  forall steps x y, hd ns = S x -> (eval (Recursion g h) steps (x :: tl ns)) = Some y ->
  eval (Recursion g h) steps ns = eval h steps (x :: y :: tl ns).

Lemma Recursion_correct_step' : forall k (g:PRFunction k) (h:PRFunction (2+k)) (ns:t nat (1+k)),
  forall steps x, hd ns = S x -> (eval (Recursion g h) steps (x :: tl ns)) = None ->
  eval (Recursion g h) steps ns = None.
\end{lstlisting}
These results rely on a number of auxiliary results, notably about the function
\lstinline+find_zero_from+.

\subsection{Examples}

For further proof of correctness, we chose some functions that typically are used as examples in
textbooks on the topic -- addition, multiplication, sign -- and some relations -- greater than,
smaller than, equals -- and showed that the usual definitions are correct.
For example, sum is defined as $R(P^1_1,C(S,P^3_2))$ (this is also used as an example in~\cite{CM20}).
Our formalisation defines \lstinline+PR_add+ as
\lstinline+Recursion (Projection aux11) (Composition Successor [Projection aux23])+, and includes
\begin{lstlisting}
Lemma add_correct : forall m n steps, eval PR_add steps [m; n] = Some (m + n).
\end{lstlisting}
(The parameters $m$ and $k$ are implicit in the constructor for projections; the proof terms are
named to correspond to the informal usage, so that \lstinline+aux11:0<1+.
Thus $P^1_1$ is represented by \lstinline+@Projection 0 1 aux11+.)

\subsection{Convergence and Uniqueness}

The next step is to show that the value by \lstinline+eval+ is unique and stable (augmenting the
number of steps can only change it from \lstinline+None+ to \lstinline+Some n+, and not conversely).

This is the first place where we have to do induction over \lstinline+PRFunction+.
The induction principle automatically generated by Coq from the type definition is not strong
enough for our purposes: the constructor for composition includes elements of type
\lstinline+PRFunction+ inside a vector argument, and these functions are not available on inductive
proofs.
We use a standard technique to overcome this limitation: we assign a depth to every element of
\lstinline+PRFunction+ (corresponding to the depth of its abstract syntax tree), and prove results
by induction over the depth of functions.
In particular, this allows us to prove the following general induction principle.
\begin{lstlisting}
Theorem PRFunction_induction : forall (P:forall (n:nat) (f:PRFunction n), Prop),
  P _ Zero -> P _ Successor ->
  (forall i j (Hp:i<j), P _ (Projection Hp)) ->
  (forall m k g fs, (forall H, P m fs[@H]) -> P k g -> P _ (Composition g fs)) ->
  (forall k g h, P _ g -> P _ h -> P (1+k) (Recursion g h)) ->
  (forall k h, P _ h -> P k (Minimization h)) ->
  forall n f, P n f.
\end{lstlisting}
(The notation \lstinline+v[@H]+ denotes the $k$-th element of vector \lstinline+v+, where
\lstinline+H+ is a proof that $k$ is smaller than the length of \lstinline+v+.)

Using this principle (and sometimes directly induction over depth), we can prove all mentioned
properties of evaluation.
We then define convergence and divergence in the natural way.
\begin{lstlisting}
Definition converges {k} (f:PRFunction k) ns y := exists steps, eval f steps ns = Some y.
Definition diverges {k} (f:PRFunction k) ns := forall steps, eval f steps ns = None.
Lemma converges_inj : forall {k} f ns y y', converges (k:=k) f ns y -> converges f ns y' -> y = y'.
Lemma converges_diverges : forall {k} f ns, (diverges (k:=k) f ns <-> forall y, ~converges f ns y).
\end{lstlisting}

Finally, we prove a number of results for establishing convergence of each class of functions.
These results are used later, when proving Turing completeness of choreographies.
For example, for recursion we have:
\begin{lstlisting}
Lemma Recursion_converges_base : forall k g h ns y,
  converges g (tl ns) y -> converges (@Recursion k g h) (0::tl ns) y.

Lemma Recursion_converges_step : forall k g h ns x y z,
  converges (@Recursion k g h) (x::ns) y ->
  converges h (x::y::ns) z -> converges (Recursion g h) (S x::ns) z.
\end{lstlisting}

Conversely, if recursion converges, then all intermediate computations must also converge.
\begin{lstlisting}
Lemma converges_Recursion_base : forall {m} (g:PRFunction m) h ns y,
  converges (Recursion g h) ns y -> hd ns = 0 -> converges g (tl ns) y.

Lemma converges_Recursion_step : forall {m} (g:PRFunction m) h ns x y,
  converges (Recursion g h) ns y -> hd ns = (S x) ->
  exists z, converges (Recursion g h) (x :: tl ns) z /\ converges h (x :: z :: tl ns) y.

Lemma converges_Recursion_full : forall {m} (g:PRFunction m) h ns y,
  converges (Recursion g h) ns y ->
  forall x, x <= hd ns -> exists z, converges (Recursion g h) (x :: tl ns) z.
\end{lstlisting}

For completeness, the formalisation also includes corresponding results for divergence; these are
currently unused.

This part of the development contains 22 definitions and 84 lemmas, with a total of 1388 lines of
code.
(This excludes some results on basic data structures that we could not find in the standard
library.)

\section{Turing Completeness of Choreographies}
\label{sec:turing}

We are now ready to show that the choreographic language is Turing complete, in the sense that every partial recursive function can be implemented as a choreography (for a suitable definition of implementation).
The construction is very similar to that in~\cite{CM20}: a significant part of the formalisation
amounted to transcribing all the relevant definitions to Coq syntax.
This contributes to confirming that the simplifications introduced in~\cite{M20itc} and the additional notions and properties (rich labels, well-formedness, etc.) introduced by our formalisation are mostly internal and aimed at simplifying metatheoretical reasoning on choreographies.

\subsection{Concrete Language}

The first step is to instantiate the parameters in the definition of \lstinline+CC+ with the right
types.
Process identifiers, values and procedure names are natural numbers.
Each process contains two variables; we use \lstinline+Bool+ for this type, and alias its elements
to \lstinline+xx+ and \lstinline+yy+ for clarity.
Expressions are an inductive type with three elements: \lstinline+this+ (evaluating to the
process's value at \lstinline+xx+), \lstinline+zero+ (evaluating to \lstinline+0+) and
\lstinline+succ_this+ (evaluating to the successor of the value at \lstinline+xx+).
Boolean expressions are a singleton type with one element \lstinline+compare+, which evaluates to
\lstinline+true+ exactly when the process's two variables store the same element.

We restrict the syntax of choreographies to mimic the operators from the original development in~\cite{CM20}, where
processes only had one storage variable.
Thus, incoming value communications are always stored at variable \lstinline+xx+.
However, in that calculus the conditional compared the value stored in two processes.
We model this as a communication whose result is stored at \lstinline+yy+, followed by a call to
\lstinline+compare+.
We define these operations as macros.
\begin{lstlisting}
Definition Send p e q := p#e --> qdollarxx.
Definition IfEq p q C1 C2 := q#this --> pdollaryy;; If p ? compare Then C1 Else C2.
\end{lstlisting}

\subsection{Encoding}

The most complex step of the formalisation is formalising the encoding of partial recursive
functions as choreographies.
This is naturally a recursive construction, but there are some challenges.
First, non-base functions need to store intermediate computation results in auxiliary processes;
second, recursion and minimisation use procedure definitions to implement loops.
The strategy in~\cite{CM20} was to use auxiliary processes sequentially: since we can statically
determine how many processes are needed from the definition of the function to encode, we can
always determine the first unused process.
The language used therein did not have the problem with recursion variables, but the same technique
apply.

The key definition is the following: a choreography $C$ implements function $f:\NN^m\to\NN$ with
input processes $p_1,\ldots,p_m$ and output process $q$ iff: for any state $s$ where
$p_1,\ldots,p_m$ contain the values $n_1,\ldots,n_m$ in their variable \lstinline+xx+, (i) if
$f(n_1,\ldots,n_m)=n$, then all executions of $C$ from $s$ terminate, and do so in a state where $q$
stores $n$ in its variable \lstinline+xx+; and (ii) if $f(n_1,\ldots,n_m)$ is undefined, then
execution of $C$ from $s$ never terminates.
This is captured in the following Coq definition.
\begin{lstlisting}
Definition implements (P:Program) {n} (f:PRFunction n) (ps:t Pid n) (q:Pid) :=
  forall (xs:t nat n) (s:State), (forall Hi, s (ps[@Hi]) xx = xs[@Hi]) ->
  (forall y, converges f xs y <-> exists s' ts P', (P,s) --[ts]-->* (P',s') /\ s' q xx = y /\ Main P' = End) /\
  (diverges f xs <-> forall s' ts P', (P,s) --[ts]-->* (P',s') -> Main P' <> End).
\end{lstlisting}

The idea is that we recursively define the set of procedure definitions needed to encode
$f:\NN^m\to\NN$, taking as parameters not only the processes $p_1,\ldots,p_m$ and $q$, but also the
indices of the first unused process and the first unused procedure.
The encoding of $f$ is a program whose set of procedure definitions is obtained by instantiating the
last two values to $\max(p_1,\ldots,p_m,q)+1$ and $0$, respectively, and whose main choreography is
\lstinline+Call 0+.
Furthermore, the encoding is done in such a way that the choreography terminates by calling the
first unused procedure (which by default is defined as the terminated choreography).
This makes it easy to ensure that procedure calls compose nicely in the recursive steps of the
construction.

We start by defining two auxiliary functions \lstinline+Pi+ and \lstinline+Gamma+, which given a
function in \lstinline+PRFunction+ return the number of processes and procedures needed to encode
it, respectively.
(Function \lstinline+Pi+ is exactly the function $\Pi$ from~\cite{CM20}.)
No results about these functions are needed -- their definition suffices to prove all needed
results, namely that no process (resp.\ procedure) higher than \lstinline+Pi f+
(resp.\ \lstinline+Gamma f+) is used when encoding \lstinline+f+.

Ironically, the most challenging part of the definition is composition.
The base cases are directly encoded by suitably adapting the definitions from~\cite{CM20}, as
well as those of recursion and minimisation.
However, the definition of composition needs to be recursive (due to the variable number of argument
functions), and working with vectors adds a layer of complexity.
As such, a number of auxiliary functions were defined to deal with composition, defining a
choreography that encodes a vector of functions all with the same inputs and returning outputs in
consecutive processes.

The recursive definition of encoding \lstinline+Encoding_rec+ is again written interactively, and
afterwards a number of lemmas prove that it behaves as expected, e.g.:
\begin{lstlisting}
Lemma Zero_Procs : forall d Hd ps q n X,
  Encoding_rec Zero d Hd ps q n X X = Send (hd ps) zero q;; Call (S X).

Lemma Recursion_Procs_g : forall k (g:PRFunction k) h d (Hd:depth (Recursion g h) < S d) ps q n,
  forall X Y, X <= Y < X + Gamma g -> let Hg := (...) in
  Encoding_rec (Recursion g h) _ Hd ps q n X Y = Encoding_rec _ _ Hg (tl ps) n (n+3) X Y.
\end{lstlisting}
where we omit the definition of the proof term \lstinline+Hg+.

Finally, we prove that encoding always returns a well-formed choreography. This is implicit
in~\cite{CM20}, but it is an essential property that should hold.
For convenience, each condition of well-formedness is proved separately, capitalising on the fact
that the encoding returns an initial choreography.
The proof follows the recursive structure of the definition of
\lstinline+Encoding_rec+, and is relatively automatic once the relevant splitting in cases is done.

\begin{lstlisting}
Lemma Encoding_WF : forall {n} (f:PRFunction n) ps q, ~In q ps -> CCP_WF (Encoding f ps q).
\end{lstlisting}

\subsection{Soundness}

Soundness of the encoding -- the property that the encoding of $f$ implements $f$ -- is proven by
analysing the execution path obtained by always reducing the first action in the choreography, and
invoking confluence.
We split the proof into a number of lemmas, stating the obvious reductions from each procedure
definition to the procedure call at its end.
We give some examples of these results.

\begin{lstlisting}
Lemma Zero_reduce : forall Defs (ps: t Pid 1) q X s,
  exists t, (Build_Program Defs (Send (hd ps) zero q;; Call X),s)
    --[t]--> (Build_Program Defs (Call X), update s q xx 0).

Lemma Recursion_reduce_0 : forall Defs X n s,
  exists t, (Build_Program Defs (Send (n + 2) zero (S n);; Call X),s)
    --[t]--> (Build_Program Defs (Call X),update s (S n) xx 0).

Lemma Recursion_reduce_1_true : forall m Defs X Y n (ps:t Pid (S m)) q s,
  s (S n) xx = s (hd ps) xx -> exists t s', (Build_Program Defs (IfEq (S n) (hd ps) (Send n this q;; Call X) (Call Y)),s)
    --[t]-->* (Build_Program Defs (Call X), s') /\ s' q xx = s n xx /\ forall p, p <> q -> s' p xx = s p xx.

Lemma Recursion_reduce_1_false : forall m Defs X Y n (ps:t Pid (S m)) q s,
  s (S n) xx <> s (hd ps) xx -> exists t, (Build_Program Defs (IfEq (S n) (hd ps) (Send n this q;; Call X) (Call Y)),s)
    --[t]-->* (Build_Program Defs (Call Y), update s (S n) yy (s (hd ps) xx)).
\end{lstlisting}
We briefly explain the lemmas about recursion.
Encoding $R(g,h)$ uses three auxiliary procedures.
The first one initializes the recursion by placing a zero on the first auxiliary process (Lemma
\lstinline+Recursion_reduce_0+), and calls the first procedure in the encoding of $g$.
The second one, placed immediately after the procedures used for encoding $g$, checks whether the
value in the auxiliary process is the value where we want to stop, and in this case places the
result in the return process (Lemma \lstinline+Recursion_reduce_1_true+).
Otherwise, it calls the first procedure in the encoding of $h$ (Lemma
\lstinline+Recursion_reduce_1_true+).
(This explanation assumes the values of \lstinline+X+ and \lstinline+Y+ to be instantiated in the
right way, but the lemmas do not depend on this.)
The third procedure, invoked when $h$ terminates, increases the value of the auxiliary process
controlling the loop (Lemma \lstinline+Recursion_reduce_2+, omitted).
All these lemmas also include characterisations of the resulting state that are needed for applying
the induction hypothesis in the main proof.

Using these results, we can prove by induction that, if $s$ is a state where $n_1,\ldots,n_m$ are
stored in the appropriate processes, (i)~if $f(n_1,\ldots,n_k)$ converges, then there is some
execution path of the encoding of $f$ from $s$ that terminates with the expected result; (ii)~using
confluence, all execution paths from $s$ must terminate in the same state; and (iii)~that if
$f(n_1,\ldots,n_k)$ diverges, then no execution of the encoding of $f$ from $s$ terminates.
These three results are combined in a single theorem, stating that the default encoding of $f$
(where $n_1,\ldots,n_m$ are stored in processes $1,\ldots,m$, and the result is returned in process
$0$) is sound.

\begin{lstlisting}
Theorem encoding_sound : forall n (f:PRFunction n), implements (Encoding' f) f (vec_1_to_n n) 0.
\end{lstlisting}

This part of the development contains 28 definitions and 65 lemmas, with a total of 2352 lines of code.

\section{Discussion}
\label{sec:discussion}

We presented a formalisation of a choreographic language and proved it Turing complete.
To the best of our knowledge, this is the first time that such a task has been achieved -- the only
comparable work in progress consists of a preliminary presentation on a certified compiler from choreographies to CakeML~\cite{GA18}, which however does not deal with the major challenge of recursion (the choreography language used therein is simplistic and can only express finite behaviour)~\cite{G20}.
Moreover, we showed how formalising proofs unveiled subtle problems in definitions and can influence the development of the theory, making a case for a more systematic use of theorem provers in research in the field.
The number of choreographic languages proposed in the literature is increasing rapidly, to include features of practical value such as asynchronous communication, non-determinism, broadcast, dynamic network topologies, and more~\cite{Aetal16,GVWY17,Hetal16}. Hopefully, our work can contribute a solid foundation for the development of these features. This recalls the situation found in the field of process calculi, and indeed similar conclusions are drawn in an article that presents the formalisation of a higher-order process calculus in Coq~\cite{MS15}.

It is interesting that the proof of Turing completeness -- both construction and proofs -- still closely
follows the original theory~\cite{CM20}, despite the significant changes to the lowest layers that we had to make.
This suggests that the major formalisation challenge currently lies in the foundational work.
To the best of our knowledge, this is also the first time that Kleene's theory of partial recursive
functions has been formalised in Coq.

Our formalisation includes some design options.
The most significant one, in our opinion, is the restriction to only two labels in selections.
However, as is well known in the field of session types, this is not a serious restriction~\cite{CP10}. Labels are typically used to communicate choices based on a conditional; more complex decisions are expressed as nested conditionals, and can be communicated by sending multiple label selections.

Restricting the set of labels to two elements also has a strong impact on the formalisation of
realisability~\cite{BBO12,CHY12}, which we do not discuss in this article.
Choreography realisability deals with identifying sufficient conditions for a choreography to be
implementable in a distributed setting, and generating an implementation in a process calculus
automatically.
We are currently finishing formalising this construction, and this process heavily relies on the set
of labels being fixed.
An important result is that any choreography can be amended into a realisable one, so that in
particular our Turing-completeness result immediately implies Turing-completeness of the process
calculus used for implementations.

We aimed at making our development reusable, so that it can readily be extended to more expressive
choreographic languages.
In the future, we plan to look at interesting extensions (such as those mentioned above) and explore how easy it is to extend the current formalisation to those frameworks.
We conjecture that this will prove much simpler than the current
effort, thanks to the structures established in this work.

The full development can be downloaded at~\cite{CMP21-source}.

\begin{anonsuppress}
\begin{acks}
	This work was partially supported by Villum Fonden, grant no.\ 29518.
\end{acks}
\end{anonsuppress}

\bibliography{biblio.bib}

%%% -*-BibTeX-*-
%%% Do NOT edit. File created by BibTeX with style
%%% ACM-Reference-Format-Journals [18-Jan-2012].

\begin{thebibliography}{31}

%%% ====================================================================
%%% NOTE TO THE USER: you can override these defaults by providing
%%% customized versions of any of these macros before the \bibliography
%%% command.  Each of them MUST provide its own final punctuation,
%%% except for \shownote{}, \showDOI{}, and \showURL{}.  The latter two
%%% do not use final punctuation, in order to avoid confusing it with
%%% the Web address.
%%%
%%% To suppress output of a particular field, define its macro to expand
%%% to an empty string, or better, \unskip, like this:
%%%
%%% \newcommand{\showDOI}[1]{\unskip}   % LaTeX syntax
%%%
%%% \def \showDOI #1{\unskip}           % plain TeX syntax
%%%
%%% ====================================================================

\ifx \showCODEN    \undefined \def \showCODEN     #1{\unskip}     \fi
\ifx \showDOI      \undefined \def \showDOI       #1{#1}\fi
\ifx \showISBNx    \undefined \def \showISBNx     #1{\unskip}     \fi
\ifx \showISBNxiii \undefined \def \showISBNxiii  #1{\unskip}     \fi
\ifx \showISSN     \undefined \def \showISSN      #1{\unskip}     \fi
\ifx \showLCCN     \undefined \def \showLCCN      #1{\unskip}     \fi
\ifx \shownote     \undefined \def \shownote      #1{#1}          \fi
\ifx \showarticletitle \undefined \def \showarticletitle #1{#1}   \fi
\ifx \showURL      \undefined \def \showURL       {\relax}        \fi
% The following commands are used for tagged output and should be
% invisible to TeX
\providecommand\bibfield[2]{#2}
\providecommand\bibinfo[2]{#2}
\providecommand\natexlab[1]{#1}
\providecommand\showeprint[2][]{arXiv:#2}

\bibitem[\protect\citeauthoryear{Albert and Lanese}{Albert and Lanese}{2016}]%
        {forte2016}
\bibfield{editor}{\bibinfo{person}{Elvira Albert} {and} \bibinfo{person}{Ivan
  Lanese}} (Eds.). \bibinfo{year}{2016}\natexlab{}.
\newblock \bibinfo{booktitle}{\emph{Formal Techniques for Distributed Objects,
  Components, and Systems - 36th {IFIP} {WG} 6.1 International Conference,
  {FORTE} 2016, Held as Part of the 11th International Federated Conference on
  Distributed Computing Techniques, DisCoTec 2016, Heraklion, Crete, Greece,
  June 6-9, 2016, Proceedings}}. \bibinfo{series}{Lecture Notes in Computer
  Science}, Vol.~\bibinfo{volume}{9688}. \bibinfo{publisher}{Springer}.
\newblock


\bibitem[\protect\citeauthoryear{Ancona, Bono, Bravetti, Campos, Castagna,
  Deni{\'{e}}lou, Gay, Gesbert, Giachino, Hu, Johnsen, Martins, Mascardi,
  Montesi, Neykova, Ng, Padovani, Vasconcelos, and Yoshida}{Ancona
  et~al\mbox{.}}{2016}]%
        {Aetal16}
\bibfield{author}{\bibinfo{person}{Davide Ancona}, \bibinfo{person}{Viviana
  Bono}, \bibinfo{person}{Mario Bravetti}, \bibinfo{person}{Joana Campos},
  \bibinfo{person}{Giuseppe Castagna}, \bibinfo{person}{Pierre{-}Malo
  Deni{\'{e}}lou}, \bibinfo{person}{Simon~J. Gay}, \bibinfo{person}{Nils
  Gesbert}, \bibinfo{person}{Elena Giachino}, \bibinfo{person}{Raymond Hu},
  \bibinfo{person}{Einar~Broch Johnsen}, \bibinfo{person}{Francisco Martins},
  \bibinfo{person}{Viviana Mascardi}, \bibinfo{person}{Fabrizio Montesi},
  \bibinfo{person}{Rumyana Neykova}, \bibinfo{person}{Nicholas Ng},
  \bibinfo{person}{Luca Padovani}, \bibinfo{person}{Vasco~T. Vasconcelos},
  {and} \bibinfo{person}{Nobuko Yoshida}.} \bibinfo{year}{2016}\natexlab{}.
\newblock \showarticletitle{Behavioral Types in Programming Languages}.
\newblock \bibinfo{journal}{\emph{Foundations and Trends in Programming
  Languages}} \bibinfo{volume}{3}, \bibinfo{number}{2--3}
  (\bibinfo{year}{2016}), \bibinfo{pages}{95--230}.
\newblock


\bibitem[\protect\citeauthoryear{Basu, Bultan, and Ouederni}{Basu
  et~al\mbox{.}}{2012}]%
        {BBO12}
\bibfield{author}{\bibinfo{person}{Samik Basu}, \bibinfo{person}{Tevfik
  Bultan}, {and} \bibinfo{person}{Meriem Ouederni}.}
  \bibinfo{year}{2012}\natexlab{}.
\newblock \showarticletitle{Deciding choreography realizability}. In
  \bibinfo{booktitle}{\emph{Procs.\ POPL}},
  \bibfield{editor}{\bibinfo{person}{John Field} {and} \bibinfo{person}{Michael
  Hicks}} (Eds.). \bibinfo{publisher}{{ACM}}, \bibinfo{pages}{191--202}.
\newblock
\urldef\tempurl%
\url{https://doi.org/10.1145/2103656.2103680}
\showDOI{\tempurl}


\bibitem[\protect\citeauthoryear{Caires and Pfenning}{Caires and
  Pfenning}{2010}]%
        {CP10}
\bibfield{author}{\bibinfo{person}{Lu{\'{\i}}s Caires} {and}
  \bibinfo{person}{Frank Pfenning}.} \bibinfo{year}{2010}\natexlab{}.
\newblock \showarticletitle{Session Types as Intuitionistic Linear
  Propositions}. In \bibinfo{booktitle}{\emph{Procs.\ CONCUR}}
  \emph{(\bibinfo{series}{Lecture Notes in Computer Science})},
  \bibfield{editor}{\bibinfo{person}{Paul Gastin} {and}
  \bibinfo{person}{Fran{\c{c}}ois Laroussinie}} (Eds.),
  Vol.~\bibinfo{volume}{6269}. \bibinfo{publisher}{Springer},
  \bibinfo{pages}{222--236}.
\newblock
\urldef\tempurl%
\url{https://doi.org/10.1007/978-3-642-15375-4\_16}
\showDOI{\tempurl}


\bibitem[\protect\citeauthoryear{Carbone, Honda, and Yoshida}{Carbone
  et~al\mbox{.}}{2012}]%
        {CHY12}
\bibfield{author}{\bibinfo{person}{Marco Carbone}, \bibinfo{person}{Kohei
  Honda}, {and} \bibinfo{person}{Nobuko Yoshida}.}
  \bibinfo{year}{2012}\natexlab{}.
\newblock \showarticletitle{Structured Communication-Centered Programming for
  Web Services}.
\newblock \bibinfo{journal}{\emph{{ACM} Trans.\ Program.\ Lang.\ Syst.}}
  \bibinfo{volume}{34}, \bibinfo{number}{2} (\bibinfo{year}{2012}),
  \bibinfo{pages}{8:1--8:78}.
\newblock
\urldef\tempurl%
\url{https://doi.org/10.1145/2220365.2220367}
\showDOI{\tempurl}


\bibitem[\protect\citeauthoryear{Carbone and Montesi}{Carbone and
  Montesi}{2013}]%
        {CM13}
\bibfield{author}{\bibinfo{person}{Marco Carbone} {and}
  \bibinfo{person}{Fabrizio Montesi}.} \bibinfo{year}{2013}\natexlab{}.
\newblock \showarticletitle{Deadlock-freedom-by-design: multiparty asynchronous
  global programming}. In \bibinfo{booktitle}{\emph{Procs.\ POPL}},
  \bibfield{editor}{\bibinfo{person}{Roberto Giacobazzi} {and}
  \bibinfo{person}{Radhia Cousot}} (Eds.). \bibinfo{publisher}{{ACM}},
  \bibinfo{pages}{263--274}.
\newblock
\urldef\tempurl%
\url{https://doi.org/10.1145/2429069.2429101}
\showDOI{\tempurl}


\bibitem[\protect\citeauthoryear{Cruz{-}Filipe and Montesi}{Cruz{-}Filipe and
  Montesi}{2016}]%
        {CM16}
\bibfield{author}{\bibinfo{person}{Lu{\'{\i}}s Cruz{-}Filipe} {and}
  \bibinfo{person}{Fabrizio Montesi}.} \bibinfo{year}{2016}\natexlab{}.
\newblock \showarticletitle{Choreographies in Practice}, See \citeN{forte2016},
  \bibinfo{pages}{114--123}.
\newblock
\urldef\tempurl%
\url{https://doi.org/10.1007/978-3-319-39570-8\_8}
\showDOI{\tempurl}


\bibitem[\protect\citeauthoryear{Cruz{-}Filipe and Montesi}{Cruz{-}Filipe and
  Montesi}{2020}]%
        {CM20}
\bibfield{author}{\bibinfo{person}{Lu{\'{\i}}s Cruz{-}Filipe} {and}
  \bibinfo{person}{Fabrizio Montesi}.} \bibinfo{year}{2020}\natexlab{}.
\newblock \showarticletitle{A core model for choreographic programming}.
\newblock \bibinfo{journal}{\emph{Theor.\ Comput.\ Sci.}}
  \bibinfo{volume}{802} (\bibinfo{year}{2020}), \bibinfo{pages}{38--66}.
\newblock
\urldef\tempurl%
\url{https://doi.org/10.1016/j.tcs.2019.07.005}
\showDOI{\tempurl}


\bibitem[\protect\citeauthoryear{Cruz-Filipe, Montesi, and
  Peressotti}{Cruz-Filipe et~al\mbox{.}}{2019}]%
        {CMP19}
\bibfield{author}{\bibinfo{person}{Lu{\'\i}s Cruz-Filipe},
  \bibinfo{person}{Fabrizio Montesi}, {and} \bibinfo{person}{Marco
  Peressotti}.} \bibinfo{year}{2019}\natexlab{}.
\newblock \showarticletitle{Choreographies in Coq}. In
  \bibinfo{booktitle}{\emph{TYPES 2019, Abstracts}}.
\newblock
\newblock
\shownote{Extended abstract.}


\bibitem[\protect\citeauthoryear{Cruz-Filipe, Montesi, and
  Peressotti}{Cruz-Filipe et~al\mbox{.}}{2021}]%
        {CMP21-source}
\bibfield{author}{\bibinfo{person}{Luís Cruz-Filipe},
  \bibinfo{person}{Fabrizio Montesi}, {and} \bibinfo{person}{Marco
  Peressotti}.} \bibinfo{year}{2021}\natexlab{}.
\newblock \bibinfo{title}{{A Formalisation of a Turing-Complete Choreographic
  Language in Coq}}.
\newblock
\newblock
\urldef\tempurl%
\url{https://doi.org/10.5281/zenodo.4479102}
\showDOI{\tempurl}


\bibitem[\protect\citeauthoryear{Dalla~Preda, Gabbrielli, Giallorenzo, Lanese,
  and Mauro}{Dalla~Preda et~al\mbox{.}}{2017}]%
        {DGGLM17}
\bibfield{author}{\bibinfo{person}{Mila Dalla~Preda}, \bibinfo{person}{Maurizio
  Gabbrielli}, \bibinfo{person}{Saverio Giallorenzo}, \bibinfo{person}{Ivan
  Lanese}, {and} \bibinfo{person}{Jacopo Mauro}.}
  \bibinfo{year}{2017}\natexlab{}.
\newblock \showarticletitle{Dynamic Choreographies: Theory And Implementation}.
\newblock \bibinfo{journal}{\emph{Log.\ Methods Comput.\ Sci.}}
  \bibinfo{volume}{13}, \bibinfo{number}{2} (\bibinfo{year}{2017}).
\newblock
\urldef\tempurl%
\url{https://doi.org/10.23638/LMCS-13(2:1)2017}
\showDOI{\tempurl}


\bibitem[\protect\citeauthoryear{Gay, Vasconcelos, Wadler, and Yoshida}{Gay
  et~al\mbox{.}}{2017}]%
        {GVWY17}
\bibfield{author}{\bibinfo{person}{Simon~J. Gay}, \bibinfo{person}{Vasco~T.
  Vasconcelos}, \bibinfo{person}{Philip Wadler}, {and} \bibinfo{person}{Nobuko
  Yoshida}.} \bibinfo{year}{2017}\natexlab{}.
\newblock \showarticletitle{Theory and Applications of Behavioural Types
  (Dagstuhl Seminar 17051)}.
\newblock \bibinfo{journal}{\emph{Dagstuhl Reports}} \bibinfo{volume}{7},
  \bibinfo{number}{1} (\bibinfo{year}{2017}), \bibinfo{pages}{158--189}.
\newblock
\urldef\tempurl%
\url{https://doi.org/10.4230/DagRep.7.1.158}
\showDOI{\tempurl}


\bibitem[\protect\citeauthoryear{Giallorenzo, Lanese, and Russo}{Giallorenzo
  et~al\mbox{.}}{2018}]%
        {GLR18}
\bibfield{author}{\bibinfo{person}{Saverio Giallorenzo}, \bibinfo{person}{Ivan
  Lanese}, {and} \bibinfo{person}{Daniel Russo}.}
  \bibinfo{year}{2018}\natexlab{}.
\newblock \showarticletitle{ChIP: {A} Choreographic Integration Process}. In
  \bibinfo{booktitle}{\emph{Procs.\ OTM, part II}}
  \emph{(\bibinfo{series}{Lecture Notes in Computer Science})},
  \bibfield{editor}{\bibinfo{person}{Herv{\'{e}} Panetto},
  \bibinfo{person}{Christophe Debruyne}, \bibinfo{person}{Henderik~A. Proper},
  \bibinfo{person}{Claudio~Agostino Ardagna}, \bibinfo{person}{Dumitru Roman},
  {and} \bibinfo{person}{Robert Meersman}} (Eds.),
  Vol.~\bibinfo{volume}{11230}. \bibinfo{publisher}{Springer},
  \bibinfo{pages}{22--40}.
\newblock
\urldef\tempurl%
\url{https://doi.org/10.1007/978-3-030-02671-4\_2}
\showDOI{\tempurl}


\bibitem[\protect\citeauthoryear{Giallorenzo, Montesi, and
  Peressotti}{Giallorenzo et~al\mbox{.}}{2020}]%
        {GMP20}
\bibfield{author}{\bibinfo{person}{Saverio Giallorenzo},
  \bibinfo{person}{Fabrizio Montesi}, {and} \bibinfo{person}{Marco
  Peressotti}.} \bibinfo{year}{2020}\natexlab{}.
\newblock \showarticletitle{Choreographies as Objects}.
\newblock \bibinfo{journal}{\emph{CoRR}}  \bibinfo{volume}{abs/2005.09520}
  (\bibinfo{year}{2020}).
\newblock
\urldef\tempurl%
\url{https://arxiv.org/abs/2005.09520}
\showURL{%
\tempurl}


\bibitem[\protect\citeauthoryear{Gomez-Londono and Aman~Pohjola}{Gomez-Londono
  and Aman~Pohjola}{2018}]%
        {GA18}
\bibfield{author}{\bibinfo{person}{Alejandro Gomez-Londono} {and}
  \bibinfo{person}{Johannes Aman~Pohjola}.} \bibinfo{year}{2018}\natexlab{}.
\newblock \showarticletitle{Connecting Choreography Languages With Verified
  Stacks}. In \bibinfo{booktitle}{\emph{Procs.\ of the Nordic Workshop on
  Programming Theory}}. \bibinfo{pages}{31--33}.
\newblock
\urldef\tempurl%
\url{http://hdl.handle.net/102.100.100/86327?index=1}
\showURL{%
\tempurl}


\bibitem[\protect\citeauthoryear{Honda, Yoshida, and Carbone}{Honda
  et~al\mbox{.}}{2016}]%
        {HYC16}
\bibfield{author}{\bibinfo{person}{Kohei Honda}, \bibinfo{person}{Nobuko
  Yoshida}, {and} \bibinfo{person}{Marco Carbone}.}
  \bibinfo{year}{2016}\natexlab{}.
\newblock \showarticletitle{Multiparty Asynchronous Session Types}.
\newblock \bibinfo{journal}{\emph{J. {ACM}}} \bibinfo{volume}{63},
  \bibinfo{number}{1} (\bibinfo{year}{2016}), \bibinfo{pages}{9}.
\newblock
\urldef\tempurl%
\url{https://doi.org/10.1145/2827695}
\showDOI{\tempurl}
\newblock
\shownote{Also: POPL, pages 273--284, 2008.}


\bibitem[\protect\citeauthoryear{H{\"{u}}ttel, Lanese, Vasconcelos, Caires,
  Carbone, Deni{\'{e}}lou, Mostrous, Padovani, Ravara, Tuosto, Vieira, and
  Zavattaro}{H{\"{u}}ttel et~al\mbox{.}}{2016}]%
        {Hetal16}
\bibfield{author}{\bibinfo{person}{Hans H{\"{u}}ttel}, \bibinfo{person}{Ivan
  Lanese}, \bibinfo{person}{Vasco~T. Vasconcelos}, \bibinfo{person}{Lu{\'{\i}}s
  Caires}, \bibinfo{person}{Marco Carbone}, \bibinfo{person}{Pierre{-}Malo
  Deni{\'{e}}lou}, \bibinfo{person}{Dimitris Mostrous}, \bibinfo{person}{Luca
  Padovani}, \bibinfo{person}{Ant{\'{o}}nio Ravara}, \bibinfo{person}{Emilio
  Tuosto}, \bibinfo{person}{Hugo~Torres Vieira}, {and}
  \bibinfo{person}{Gianluigi Zavattaro}.} \bibinfo{year}{2016}\natexlab{}.
\newblock \showarticletitle{Foundations of Session Types and Behavioural
  Contracts}.
\newblock \bibinfo{journal}{\emph{{ACM} Comput.\ Surv.}} \bibinfo{volume}{49},
  \bibinfo{number}{1} (\bibinfo{year}{2016}), \bibinfo{pages}{3:1--3:36}.
\newblock
\urldef\tempurl%
\url{https://doi.org/10.1145/2873052}
\showDOI{\tempurl}


\bibitem[\protect\citeauthoryear{{Intl. Telecommunication Union}}{{Intl.
  Telecommunication Union}}{1996}]%
        {msc}
\bibfield{author}{\bibinfo{person}{{Intl. Telecommunication Union}}.}
  \bibinfo{year}{1996}\natexlab{}.
\newblock \bibinfo{title}{Recommendation \mbox{Z.120}: {Message Sequence
  Chart}}.
\newblock
\newblock


\bibitem[\protect\citeauthoryear{Kleene}{Kleene}{1952}]%
        {Kleene52}
\bibfield{author}{\bibinfo{person}{Stephen~Cole Kleene}.}
  \bibinfo{year}{1952}\natexlab{}.
\newblock \bibinfo{booktitle}{\emph{Introduction to Metamathematics}}.
  Vol.~\bibinfo{volume}{1}.
\newblock \bibinfo{publisher}{North-Holland Publishing Co.}
\newblock


\bibitem[\protect\citeauthoryear{Lluch{-}Lafuente, Nielson, and
  Nielson}{Lluch{-}Lafuente et~al\mbox{.}}{2015}]%
        {LN15}
\bibfield{author}{\bibinfo{person}{Alberto Lluch{-}Lafuente},
  \bibinfo{person}{Flemming Nielson}, {and} \bibinfo{person}{Hanne~Riis
  Nielson}.} \bibinfo{year}{2015}\natexlab{}.
\newblock \showarticletitle{Discretionary Information Flow Control for
  Interaction-Oriented Specifications}. In \bibinfo{booktitle}{\emph{Logic,
  Rewriting, and Concurrency}} \emph{(\bibinfo{series}{Lecture Notes in
  Computer Science})}, \bibfield{editor}{\bibinfo{person}{Narciso
  Mart{\'{\i}}{-}Oliet}, \bibinfo{person}{Peter~Csaba {\"{O}}lveczky}, {and}
  \bibinfo{person}{Carolyn~L. Talcott}} (Eds.), Vol.~\bibinfo{volume}{9200}.
  \bibinfo{publisher}{Springer}, \bibinfo{pages}{427--450}.
\newblock
\urldef\tempurl%
\url{https://doi.org/10.1007/978-3-319-23165-5\_20}
\showDOI{\tempurl}


\bibitem[\protect\citeauthoryear{Londo{\~n}o}{Londo{\~n}o}{2020}]%
        {G20}
\bibfield{author}{\bibinfo{person}{Alejandro~G{\'o}mez Londo{\~n}o}.}
  \bibinfo{year}{2020}\natexlab{}.
\newblock \bibinfo{title}{Choreographies and Cost Semantics for Reliable
  Communicating Systems}.
\newblock
\newblock


\bibitem[\protect\citeauthoryear{L{\'{o}}pez and Heussen}{L{\'{o}}pez and
  Heussen}{2017}]%
        {LH17}
\bibfield{author}{\bibinfo{person}{Hugo~A. L{\'{o}}pez} {and}
  \bibinfo{person}{Kai Heussen}.} \bibinfo{year}{2017}\natexlab{}.
\newblock \showarticletitle{Choreographing cyber-physical distributed control
  systems for the energy sector}. In \bibinfo{booktitle}{\emph{Procs.\ SAC}},
  \bibfield{editor}{\bibinfo{person}{Ahmed Seffah}, \bibinfo{person}{Birgit
  Penzenstadler}, \bibinfo{person}{Carina Alves}, {and} \bibinfo{person}{Xin
  Peng}} (Eds.). \bibinfo{publisher}{{ACM}}, \bibinfo{pages}{437--443}.
\newblock
\urldef\tempurl%
\url{https://doi.org/10.1145/3019612.3019656}
\showDOI{\tempurl}


\bibitem[\protect\citeauthoryear{L{\'{o}}pez, Nielson, and Nielson}{L{\'{o}}pez
  et~al\mbox{.}}{2016}]%
        {LNN16}
\bibfield{author}{\bibinfo{person}{Hugo~A. L{\'{o}}pez},
  \bibinfo{person}{Flemming Nielson}, {and} \bibinfo{person}{Hanne~Riis
  Nielson}.} \bibinfo{year}{2016}\natexlab{}.
\newblock \showarticletitle{Enforcing Availability in Failure-Aware
  Communicating Systems}, See \citeN{forte2016}, \bibinfo{pages}{195--211}.
\newblock
\urldef\tempurl%
\url{https://doi.org/10.1007/978-3-319-39570-8\_13}
\showDOI{\tempurl}


\bibitem[\protect\citeauthoryear{Maksimovic and Schmitt}{Maksimovic and
  Schmitt}{2015}]%
        {MS15}
\bibfield{author}{\bibinfo{person}{Petar Maksimovic} {and}
  \bibinfo{person}{Alan Schmitt}.} \bibinfo{year}{2015}\natexlab{}.
\newblock \showarticletitle{HOCore in Coq}. In
  \bibinfo{booktitle}{\emph{Interactive Theorem Proving - 6th International
  Conference, {ITP} 2015, Nanjing, China, August 24-27, 2015, Proceedings}}
  \emph{(\bibinfo{series}{Lecture Notes in Computer Science})},
  \bibfield{editor}{\bibinfo{person}{Christian Urban} {and}
  \bibinfo{person}{Xingyuan Zhang}} (Eds.), Vol.~\bibinfo{volume}{9236}.
  \bibinfo{publisher}{Springer}, \bibinfo{pages}{278--293}.
\newblock
\urldef\tempurl%
\url{https://doi.org/10.1007/978-3-319-22102-1\_19}
\showDOI{\tempurl}


\bibitem[\protect\citeauthoryear{Montesi}{Montesi}{2013}]%
        {M13p}
\bibfield{author}{\bibinfo{person}{Fabrizio Montesi}.}
  \bibinfo{year}{2013}\natexlab{}.
\newblock \emph{\bibinfo{title}{Choreographic Programming}}.
\newblock {Ph.{D}. Thesis}. \bibinfo{school}{IT University of Copenhagen}.
\newblock
\newblock
\shownote{\href{http://www.fabriziomontesi.com/files/choreographic\_programming.pdf}{http://www.fabriziomontesi.com/files/choreographic\_programming.pdf}.}


\bibitem[\protect\citeauthoryear{Montesi}{Montesi}{2020}]%
        {M20itc}
\bibfield{author}{\bibinfo{person}{Fabrizio Montesi}.}
  \bibinfo{year}{2020}\natexlab{}.
\newblock \bibinfo{title}{{Introduction to Choreographies}}.
  (\bibinfo{year}{2020}).
\newblock
\newblock
\shownote{Accepted for publication by Cambridge University Press.}


\bibitem[\protect\citeauthoryear{Needham and Schroeder}{Needham and
  Schroeder}{1978}]%
        {NS78}
\bibfield{author}{\bibinfo{person}{Roger~M. Needham} {and}
  \bibinfo{person}{Michael~D. Schroeder}.} \bibinfo{year}{1978}\natexlab{}.
\newblock \showarticletitle{Using Encryption for Authentication in Large
  Networks of Computers}.
\newblock \bibinfo{journal}{\emph{Commun.\ {ACM}}} \bibinfo{volume}{21},
  \bibinfo{number}{12} (\bibinfo{year}{1978}), \bibinfo{pages}{993--999}.
\newblock
\urldef\tempurl%
\url{https://doi.org/10.1145/359657.359659}
\showDOI{\tempurl}


\bibitem[\protect\citeauthoryear{{O}bject~{M}anagement
  {G}roup}{{O}bject~{M}anagement {G}roup}{2011}]%
        {bpmn}
\bibfield{author}{\bibinfo{person}{{O}bject~{M}anagement {G}roup}.}
  \bibinfo{year}{2011}\natexlab{}.
\newblock \bibinfo{title}{{B}usiness {P}rocess {M}odel and {N}otation}.
\newblock
  \bibinfo{howpublished}{\href{http://www.omg.org/spec/BPMN/2.0/}{http://www.omg.org/spec/BPMN/2.0/}}.
\newblock


\bibitem[\protect\citeauthoryear{Scalas and Yoshida}{Scalas and
  Yoshida}{2019}]%
        {SY19}
\bibfield{author}{\bibinfo{person}{Alceste Scalas} {and}
  \bibinfo{person}{Nobuko Yoshida}.} \bibinfo{year}{2019}\natexlab{}.
\newblock \showarticletitle{Less is more: multiparty session types revisited}.
\newblock \bibinfo{journal}{\emph{Proc.\ {ACM} Program.\ Lang.}}
  \bibinfo{volume}{3}, \bibinfo{number}{{POPL}} (\bibinfo{year}{2019}),
  \bibinfo{pages}{30:1--30:29}.
\newblock
\urldef\tempurl%
\url{https://doi.org/10.1145/3290343}
\showDOI{\tempurl}


\bibitem[\protect\citeauthoryear{Turing}{Turing}{1937}]%
        {Turing36}
\bibfield{author}{\bibinfo{person}{Alan~M. Turing}.}
  \bibinfo{year}{1937}\natexlab{}.
\newblock \showarticletitle{Computability and {$\lambda$}-Definability}.
\newblock \bibinfo{journal}{\emph{J. Symb.\ Log.}} \bibinfo{volume}{2},
  \bibinfo{number}{4} (\bibinfo{year}{1937}), \bibinfo{pages}{153--163}.
\newblock
\urldef\tempurl%
\url{https://doi.org/10.2307/2268280}
\showDOI{\tempurl}


\bibitem[\protect\citeauthoryear{{W3C}}{{W3C}}{2004}]%
        {wscdl}
\bibfield{author}{\bibinfo{person}{{W3C}}.} \bibinfo{year}{2004}\natexlab{}.
\newblock \bibinfo{title}{{WS Choreography Description Language}}.
\newblock
  \bibinfo{howpublished}{\href{http://www.w3.org/TR/ws-cdl-10/}{http://www.w3.org/TR/ws-cdl-10/}}.
\newblock


\end{thebibliography}

\end{document}